\documentclass[aps,prd,twocolumn,superscriptaddress,amsmath,amssymb,amsthm,nofootinbib, preprintnumbers]{revtex4-2}
\usepackage{url}
\usepackage[colorlinks=true,
    linkcolor=blue,
    citecolor=blue,
    urlcolor=blue]{hyperref}
 \usepackage{amsmath}
\usepackage{graphicx}
\usepackage{epstopdf}
\usepackage{physics}
\usepackage{float}
\usepackage{hyperref}
\usepackage{color}
\usepackage[T1]{fontenc}
\usepackage[utf8]{inputenc}
\usepackage[toc,page]{appendix}
\usepackage[usenames,dvipsnames]{xcolor}
\usepackage[normalem]{ulem}
\usepackage{enumitem}

\usepackage{tikz}
\usetikzlibrary{shapes,positioning,shadows,graphs.standard,automata,arrows}
\usepackage{pgfplots}
\pgfplotsset{compat=newest, width=2.669cm, height=2.669cm, scale only axis=true,enlargelimits=false}
\pgfplotsset{tick label style={font=\tiny}}
\pgfplotsset{every major tick/.append style={major tick length=3pt}}
\pgfplotsset{every minor tick/.append style={minor tick length=1.5pt}}
\usepgfplotslibrary{groupplots}
\usepackage{caption}
\usepackage[font=small,labelfont=bf,justification=justified,format=plain]{caption}

\usepackage[export]{adjustbox}

\providecommand{\renewoperator}[3]{%
\renewcommand*{#1}{\mathop{#2}#3}}

\makeatletter
\providecommand*{\diff}%
{\@ifnextchar^{\DIfF}{\DIfF^{}}}
\def\DIfF^#1{%
\mathop{\mathrm{\mathstrut d}}%
\nolimits^{#1}\gobblespace}
\def\gobblespace{%
\futurelet\diffarg\opspace}
\def\opspace{%
\let\DiffSpace\!%
\ifx\diffarg(%
\let\DiffSpace\relax
\else
\ifx\diffarg[%
\let\DiffSpace\relax
\else
\ifx\diffarg\{%
\let\DiffSpace\relax
\fi\fi\fi\DiffSpace}

\renewoperator{\Re}{\mathrm{Re}}{\nolimits}
\renewoperator{\Im}{\mathrm{Im}}{\nolimits}

\usepackage{bm}

\newcommand{\be}{\begin{equation}}
\newcommand{\ee}{\end{equation}}
\newcommand{\ba}{\begin{eqnarray}}
\newcommand{\ea}{\end{eqnarray}}

\newcommand{\beq}{\begin{equation}}
\newcommand{\eeq}{\end{equation}}
\newcommand{\beqa}{\begin{eqnarray}}
\newcommand{\eeqa}{\end{eqnarray}}

\usepackage{appendix}

\usepackage{xcolor}
\usepackage{scalerel}
\usepackage{tikz}
\usetikzlibrary{svg.path}
\definecolor{darkmagenta}{HTML}{8B008B}
\definecolor{orcidlogocol}{HTML}{A6CE39}
\tikzset{
  orcidlogo/.pic={
    \fill[orcidlogocol] svg{M256,128c0,70.7-57.3,128-128,128C57.3,256,0,198.7,0,128C0,57.3,57.3,0,128,0C198.7,0,256,57.3,256,128z};
    \fill[white] svg{M86.3,186.2H70.9V79.1h15.4v48.4V186.2z}
                 svg{M108.9,79.1h41.6c39.6,0,57,28.3,57,53.6c0,27.5-21.5,53.6-56.8,53.6h-41.8V79.1z M124.3,172.4h24.5c34.9,0,42.9-26.5,42.9-39.7c0-21.5-13.7-39.7-43.7-39.7h-23.7V172.4z}
                 svg{M88.7,56.8c0,5.5-4.5,10.1-10.1,10.1c-5.6,0-10.1-4.6-10.1-10.1c0-5.6,4.5-10.1,10.1-10.1C84.2,46.7,88.7,51.3,88.7,56.8z};
  }
}

\newcommand\orcidicon[1]{\href{https://orcid.org/#1}{\mbox{\scalerel*{
\begin{tikzpicture}[yscale=-1, scale=0.05, transform shape]
\pic{orcidlogo};
\end{tikzpicture}
}{|}}}}

\makeatletter
\let\oldcontentsline\contentsline
\renewcommand{\contentsline}[4]{%
  \oldcontentsline{#1}{#2}{}{#4}%
}
\makeatother

\begin{document}

\title{From Bertotti–Robinson to Vacuum: New Exact Solutions in General Relativity via Harrison and Inversion Symmetries}

\author{Jos{\'e} Barrientos\,$^{\orcidicon{0000-0003-3445-8151}}$}
\email{jbarrientos@academicos.uta.cl}
\affiliation{Institute of Mathematics of the Czech Academy of Sciences, {\v Z}itn{\'a} 25, 115 67 Prague 1, Czech Republic}
\affiliation{Vicerrector\'ia de Investigaci\'on y Postgrado, Universidad de La Serena, La Serena 1700000, Chile}

\author{Adolfo Cisterna\,$^{\orcidicon{0000-0002-6949-7882}}$}
\email{adolfo.cisterna@mff.cuni.cz}
\affiliation{Sede Esmeralda, Universidad de Tarapac{\'a}, Avenida Luis Emilio Recabarren 2477, Iquique, Chile}
\affiliation{Institute of Theoretical Physics, Faculty of Mathematics and Physics,
Charles University, V Hole{\v s}ovick\'ach 2, 180 00 Prague 8, Czech Republic}

\author{Amaro Díaz\,$^{\orcidicon{0009-0000-0262-1204}}$}
\email{amdiaz2022@udec.cl}
\affiliation{Departamento de F\'isica, Universidad de Concepci\'on,
Casilla, 160-C, Concepci\'on, Chile}

\author{Keanu M{\"u}ller\,$^{\orcidicon{0009-0007-1048-1847}}$}
\email{keanumuller2016@udec.cl}
\affiliation{Departamento de F\'isica, Universidad de Concepci\'on,
Casilla, 160-C, Concepci\'on, Chile}

\begin{abstract}
We construct new vacuum solutions of the Einstein equations generated from electrovacuum configurations embedded in external electromagnetic backgrounds. Starting from accelerating Bertotti–Robinson black holes, we exploit two independent symmetries of the electrovacuum: a Melvin–Bonnor-type magnetization and a magnetic Inversion. In both constructions, the external electromagnetic field can be removed while still leaving a non-trivial gravitational backreaction in the metric, yielding new accelerating vacuum spacetimes of Petrov type I.
In the static, non-accelerating limit, the magnetized Bertotti–Robinson–Schwarzschild case reproduces known results, while the Inversion symmetry produces a genuinely new vacuum configuration, a two-parameter extension of the Schwarzschild–Levi-Civita geometry. These constructions provide a systematic method for generating algebraically general vacuum geometries and illustrate how electromagnetic embeddings can induce non-trivial vacuum metrics in General Relativity. The main geometrical properties of these spacetimes are analyzed.
Additionally, we present two further results: a stationary generalization of these vacuum geometries and two new static vacuum configurations obtained by applying the same symmetries to the Alekseev–García black hole seed.

\end{abstract}

\maketitle

\section{Introduction}

Exact solutions occupy a privileged role in General Relativity (GR), as they provide precise spacetimes in which the non-linear Einstein equations are satisfied without approximation \cite{Stephani:2003tm}. They serve as benchmarks for testing physical intuition, numerical methods, and perturbative approaches, and often reveal phenomena that would otherwise remain hidden. 

The Plebański–Demiański family constitutes one of the most general exact electrovacuum solutions of the Einstein–Maxwell equations \cite{Plebanski:1976gy}. It unifies a large set of physically relevant spacetimes—including Kerr \cite{Kerr:1963ud}, Kerr–Newman \cite{Newman:1965my}, Taub–NUT \cite{Taub:1950ez,Newman:1963yy}, the C-metric \cite{Weyl:1917gp,LeviCivita1919,Kinnersley:1970zw}, and their extension with cosmological constant —within a single parametric framework.

A defining feature of these geometries is their Petrov type D character, for which the Weyl tensor possesses two repeated principal null directions. In this family, the electromagnetic field is aligned with these principal null directions, a property that considerably simplifies the field equations and enables the separability of geodesic and wave dynamics \cite{Carter:1968ks}. Consequently, the Plebański–Demiański metrics play a central role in the classification and interpretation of black hole solutions in electrovacuum GR.
Within this broad class, the Kerr solution is the most astrophysically relevant sector, describing the exterior gravitational field of an isolated rotating compact object. In asymptotically flat vacuum spacetimes, it is singled out by the black hole uniqueness theorems \cite{Carter:1971zc,Wald:1971iw}, which state that a stationary, axisymmetric black hole is completely characterized by its mass and angular momentum (and electromagnetic charges in the electrovacuum case). Its special status also makes it a natural reference background against which more general or non-isolated configurations are compared.

For these reasons, it is natural to seek extensions beyond the Plebański–Demiański class on theoretical grounds. One possible direction is to investigate more general algebraic families, in particular Petrov type I geometries, which are not restricted by the special alignment properties underlying many known solutions. A complementary approach is to retain algebraic specialty while enlarging the matter sector, for example, by considering electrovacuum configurations in which the electromagnetic field is not aligned with the principal null directions of the Weyl tensor. Both strategies aim to expand the set of exact solutions and clarify the extent to which current classification results depend on the assumptions embedded in the Plebański–Demiański framework.

The first direction has been extensively explored in recent years. In particular, a broad Petrov type I generalization of the Plebański–Demiański class has been identified \cite{Barrientos:2023tqb,Barrientos:2023dlf}, arising from the non-trivial interplay between accelerating spacetimes and NUT charge and electromagnetic charges. The origin of these solutions has subsequently been clarified by relating them to the general construction of exact binary configurations \cite{Astorino:2023uim}. In an appropriate large-mass limit, such systems naturally connect with accelerating geometries, thereby providing a coherent interpretation of these more general spacetimes.

In addition, algebraically general spacetimes can be constructed from any Plebański–Demiański seed by embedding the configuration in external electromagnetic and/or vortex-like backgrounds \cite{Ernst:1976bsr,Contopoulos:2015wra,Barrientos:2024pkt}. Such embeddings typically break the alignment properties that give rise to algebraic specialty, thereby promoting the geometry to Petrov type I. This mechanism provides a systematic way to generate broader families of solutions while retaining a clear relation to the original seed spacetime.

More recently, two important developments have opened a new route for exploring exact solutions in this context. The first is the construction of the Schwarzschild–Bertotti–Robinson black hole by Alekseev and García \cite{Alekseev:1996fq},\footnote{A recent charged generalization has also been provided \cite{Alekseev:2025czq}.}  which was only recently recognized to possess Petrov type D \cite{Ortaggio:2025sip}. The second is the work of Van den Bergh and Carminati \cite{VandenBergh:2020lvf}, who systematically investigated Robinson–Trautman spacetimes coupled to non-aligned electromagnetic fields. Together, these results indicate that algebraically special geometries can persist even in the presence of external electromagnetic backgrounds that are not aligned with the principal null directions.

In this respect, the formalism developed by Van den Bergh and Carminati has recently motivated the construction of a Kerr-like black hole embedded in a Bertotti–Robinson background \cite{Podolsky:2025tle}. From this framework, an apparently new Schwarzschild–Bertotti–Robinson solution has also been identified, whose precise relation to the earlier Alekseev–García geometry still remains to be clarified. Moreover, the construction yields a broader family of static metrics supported by non-aligned electromagnetic fields, which nevertheless retain a Petrov type D character \cite{Ovcharenko:2026byw}. These solutions may be interpreted, up to a mixing of rotation and acceleration that is yet to be made explicit, as an analog of the Plebański–Demiański class formulated on a Bertotti–Robinson asymptotic background.

{This Bertotti–Robinson–Plebański–Demiański family also admits further embeddings into external backgrounds of Melvin–Bonnor and/or vortex-like character. In particular, one may combine the Bertotti–Robinson and Melvin–Bonnor fields so that the overall electromagnetic configuration cancels globally, while a non-trivial gravitational backreaction persists at the level of the metric. In this way, new vacuum geometries can arise, even though their origin lies in the interaction of two electromagnetic backgrounds. The resulting spacetimes retain a general algebraic character despite the absence of a net electromagnetic field. At the non-accelerating level, this mechanism already appears to operate and has, in fact, been realized in the Schwarzschild case \cite{Astorino:2025lih}.}

In the present work, we employ two mechanisms to construct novel vacuum spacetimes of Petrov type I. First, we apply a Melvin–Bonnor-type magnetization to the recently proposed accelerating Bertotti–Robinson black holes \cite{Ovcharenko:2026byw}, thereby yielding a new family of accelerating vacuum black hole solutions. Second, we exploit an additional symmetry of the Einstein–Maxwell system, namely the Inversion symmetry, to achieve the same effect: removing the external electromagnetic field while preserving a non-trivial metric. This procedure is implemented not only for the accelerating configurations \cite{Ovcharenko:2026byw} but also at the Schwarzschild level \cite{Podolsky:2025tle}, where such a construction had not previously been identified.

The paper is organized as follows. In \autoref{sec2} we introduce the elements required for our construction: the seed spacetime, the accelerating Bertotti–Robinson configuration, and the two electrovacuum symmetries employed in our analysis, namely the magnetizing Harrison symmetry and the magnetic Inversion symmetry. In order to avoid repeated terminology, we shall refer to the latter as azimuthal Inversion. In \autoref{sec3} we present two new geometries: the Melvin–Bonnor–accelerating–Bertotti–Robinson configuration, from which our novel vacuum solutions follow, and the Levi-Civita–accelerating–Bertotti–Robinson configuration obtained via the azimuthal Inversion symmetry. We proceed to analyze several of their main geometric features. We summarize our results in \autoref{sec4}. In addition, Appendix \ref{appA} contains a vortex-like generalization of the vacuum metrics, while Appendix \ref{appB} discusses these ideas in the context of the original Alekseev–García solution, from where two extra novel vacuum spacetimes are constructed. 

\section{Preliminaries}\label{sec2}

To construct our solutions, two key ingredients must first be introduced: the seed spacetime—the family of static and accelerating metrics introduced in \cite{Ovcharenko:2026byw}—and the magnetizing Harrison and azimuthal Inversion symmetries of the Einstein–Maxwell equations. For completeness, both symmetries are presented in the Ernst-potential formalism; however, their action is also expressed in metric form, which is the formulation employed in our analysis.
In what follows, we briefly review these elements.

\subsection{The seed spacetime: An accelerating Bertotti--Robinson black hole}

The accelerating Bertotti--Robinson configuration has been recently introduced in \cite{Ovcharenko:2026byw}. It is described by the line element
\begin{equation}
\mathrm{d} s^2=\frac{1}{\Omega^2}\left[-Q \mathrm{d} \tau^2+\frac{\mathrm{d} r^2}{Q}+r^2\left(\frac{\mathrm{d} \theta^2}{P}+P \sin ^2 \theta \mathrm{d} \varphi^2\right)\right] ,\label{cmetricBR}
\end{equation}
where, 
\begin{subequations}
\begin{align}
P & =1-2 \alpha m \cos \theta+B^2 m^2 \cos ^2 \theta, \\
Q & =\left(1-B^2 m^2-\frac{2 m}{r}\right)\left(1+\left(B^2-\alpha^2\right) r^2\right), \\
\Omega^2 & =(1-\alpha r \cos \theta)^2+B^2\left[r^2\left(P-\cos ^2 \theta\right)+2 m r \cos ^2 \theta\right], 
\end{align}\label{seedfunctions}
\end{subequations}
and the gauge field 
\begin{equation}
A=\frac{1}{B}(r \Omega_{,r}-\Omega+1) \mathrm{d} \varphi. \label{magneticfield}
\end{equation}
Here we restrict attention to the purely magnetic case, since in this setting the construction of the new vacuum solution becomes possible. The relevant limiting procedures have already been analyzed in \cite{Ovcharenko:2026byw}, thereby justifying the description of the geometry as an accelerating Bertotti–Robinson spacetime. In the zero-mass limit, the solution reduces to the Bertotti–Robinson background in uniformly accelerating coordinates, whereas switching off the magnetic field recovers the standard C-metric. Conversely, vanishing acceleration leads to the Schwarzschild–Bertotti–Robinson configuration discussed in \cite{Podolsky:2025tle}, which is characterized by 
\begin{subequations}
\begin{align}
P & =1+B^2 m^2 \cos ^2 \theta, \\
Q & =\left(1-B^2 m^2-\frac{2 m}{r}\right)\left(1+B^2 r^2\right), \\
\Omega^2 & =1+B^2\left[r^2 \sin ^2 \theta+\left(2 m r+B^2 m^2 r^2\right) \cos ^2 \theta\right],
\end{align}
\end{subequations}
with the same gauge field \eqref{magneticfield}.
As usual, the acceleration is provided by a conical defect; in this case, 
\begin{equation}
  \lim_{\theta\rightarrow 0,\pi}\frac{1}{\sin\theta}\int_0^{2\pi}\sqrt{\frac{g_{\varphi\varphi}}{g_{\theta\theta}}}\mathrm{d}\varphi=2\pi(1\mp2\alpha m+B^2m^2).
\end{equation}
It is noteworthy that a regime resembling slow acceleration naturally emerges. In particular, no acceleration horizon is present when $B=\alpha$. A detailed analysis of the causal structure of this solution will be presented elsewhere, since a proper treatment of conformal infinity lies beyond the scope of the present work.

At this point, it is appropriate to clarify why, in the present work, we mostly adopt the family of solutions originating from \cite{Ovcharenko:2026byw,Podolsky:2025tle} rather than the class constructed by Alekseev and García \cite{Alekseev:1996fq}.

Initially, the Alekseev–García spacetime was first interpreted as Petrov type I \cite{Ortaggio:2018ikt}, and therefore inequivalent to the static limit of \cite{Podolsky:2025tle}, whose geometries are of Petrov type D. This issue has since been resolved: the Alekseev–García solution is in fact also of type D \cite{Ortaggio:2025sip}. Consequently, the possibility of a local equivalence between both static configurations naturally arises. However, no coordinate transformation relating the two metrics is known, and there is presently no proof of local isometry between them. 

More importantly, one can distinguish the solutions by global geometric properties that do not depend on coordinate choices.

The essential difference lies in the structure of the symmetry axis and, therefore, in the global topology of the spacetime.

The Alekseev–García solution genuinely represents a Schwarzschild black hole embedded in a Bertotti--Robinson universe. The Bertotti–Robinson background has spatial topology 
$\mathbb{R}\times S^2$, and this compact angular sector is preserved by the solution. As a consequence, the axial Killing vector possesses two disconnected fixed-point sets corresponding to the north and south poles of the $S^2$. In other words, the axis of symmetry is not a single line but consists of two distinct points separated by a sphere. This is reflected directly in the line element, which exhibits an antipodal (second) axis of symmetry.

Because the angular surfaces are compact, regularity cannot, in general, be imposed simultaneously at both poles. One of the axes, therefore, behaves as a physical strut, producing a conical defect. Additionally, this acts as an induced source (often interpreted as a mirror mass) which is not hidden behind an event horizon and consequently corresponds to a naked curvature singularity. Nevertheless, the solution preserves the true global topology of the Bertotti–Robinson background.

In contrast, the spacetime of \cite{Podolsky:2025tle} possesses a single connected symmetry axis. The north and south poles correspond merely to two directions of the same line of fixed points of the azimuthal Killing vector, and the axis is therefore 
$\mathbb{R}$-like. The angular two-surfaces are non-compact, and the spacetime is asymptotically open. The solution may be interpreted, in the static limit, as a Schwarzschild black hole immersed in an external electromagnetic field that locally resembles the Bertotti–Robinson geometry but does not share its global topology. Both families coincide in the zero–mass limit, where the Bertotti–Robinson spacetime is recovered, yet they represent different global extensions once the black hole is present.

In the present paper, we primarily employ the family of \cite{Ovcharenko:2026byw} because the presence of a single symmetry axis allows the removal of conical defects by a suitable choice of the azimuthal periodicity, thereby avoiding the unavoidable naked singularities associated with the second axis of the Alekseev–García geometry. Nevertheless, for completeness, we also construct new vacuum solutions using an Alekseev–García seed and present them in the Appendix, although a detailed geometrical analysis of that branch is beyond the scope of this work.

In what follows, we restrict ourselves to using \eqref{cmetricBR} solely as a starting point for constructing new vacuum solutions.

\subsection{The Harrison and Inversion symmetries}


It is well known that the Einstein–Maxwell theory exhibits a high degree of integrability upon reduction by spacetime isometries. In particular, when two commuting Killing vectors are present—typically those associated with stationarity and axisymmetry—the theory reduces to an effective system defined on a two-dimensional orbit space. After an appropriate Kaluza–Klein reduction and dualization of the vector fields, the resulting dynamics can be cast as a non-linear sigma model coupled to two-dimensional gravity. The dynamical degrees of freedom are encoded in the complex Ernst potentials, one of gravitational and one of electromagnetic origin, constructed from the norms and twists of the Killing vectors together with the relevant components of the gauge potential. The field equations then reduce to the Ernst equations, a pair of coupled non-linear elliptic equations defined on the two-dimensional base space.

The reduction along, for instance, $\xi=\partial_t$ and $\chi=\partial_\varphi$, is most transparently performed using a Weyl–Lewis–Papapetrou (WLP) metric, which is adapted to the symmetries generated by $\xi$ and $\chi$ and manifestly encodes circularity. This line element, however, is not unique, since the timelike and spacelike character of $\xi$ and $\chi$ can be interchanged through a discrete double Wick rotation, $t\rightarrow i\varphi$ and $\varphi\rightarrow it$. When working with this alternative form of the WLP metric—hereafter referred to as the magnetic WLP—the Einstein–Maxwell field equations take the Ernst form
\begin{align}
(\operatorname{Re} \mathcal{E}_0+\Phi_0 \bar{\Phi}_0) \nabla^2 \mathcal{E}_0 & =\nabla \mathcal{E}_0 \cdot(\nabla \mathcal{E}_0+2 \bar{\Phi}_0 \nabla \Phi_0), \\
(\operatorname{Re} \mathcal{E}_0+\Phi_0 \bar{\Phi}_0) \nabla^2 \Phi_0 & =\nabla \Phi_0 \cdot(\nabla \mathcal{E}_0+2 \bar{\Phi}_0 \nabla \Phi_0), \label{ernstequations}
\end{align}
where the complex Ernst potentials $\{\mathcal{E}_0,\Phi_0\}$ 
\begin{equation}
\mathcal{E}_0:=-f_0-\Phi_0 \bar{\Phi}_0, \quad \Phi_0:=A_{\varphi_0}, \label{ernstpotentials}
\end{equation}
are defined in terms of the functions characterizing the magnetic WLP  configuration 
\begin{equation}
\mathrm{d}s_0^2=f_0\mathrm{d}\varphi^2+\frac{1}{f_0}[e^{2\gamma_0}(\mathrm{d}\rho^2+\mathrm{d}z^2)-\rho^2\mathrm{d}t^2],\quad A_0=A_{\varphi_0} \mathrm{d}\varphi.
\end{equation}
Recall that the functions $\{f_0,\gamma_0,A_{\varphi_0}\}$ are functions of the non-Killing coordinates $\{\rho,z\}$. Here, the subscript $0$ denotes the seed geometry in what follows.
Note that we have presented a simplified form of \eqref{ernstpotentials}, since we restrict our attention to a static magnetic WLP configuration, for which no twist components arise. Moreover, in line with the goals of the present work, we consider only a magnetic component of the gauge field.

This formulation of the Einstein–Maxwell system is particularly useful for two main reasons. First, the Ernst equations are defined on the two-dimensional orbit space, a flat Euclidean space with cylindrical symmetry, thereby greatly simplifying the direct analysis of the field equations. Indeed, for example, the Alekseev–Garc\'ia black hole was obtained using the monodromy data technique developed in Refs. \cite{Alekseev1,Alekseev2,Alekseev3}, which is based on the direct integration of the Ernst equations. Second, the sigma model arising from the reduction to the two-dimensional base space is integrable and possesses an infinite-dimensional symmetry group, known as the Geroch group \cite{Geroch:1970nt}. Since the Ernst equations are simply the field equations of this two-dimensional sigma model expressed in terms of the Ernst potentials, they are invariant under the action of the Geroch group. These global symmetries therefore provide a powerful tool for constructing new stationary and axisymmetric solutions of the Einstein–Maxwell equations and have been widely exploited through transformations such as those of Ehlers \cite{Ehlers:1959aug}, Harrison \cite{Harrison}, and others \cite{Kinners,Kramer}.

In the present work, we employ two of these symmetries: a magnetizing Harrison transformation \cite{Harrison} and the symmetry we here term azimuthal Inversion. The latter designation highlights that the transformation acts on a magnetic WLP configuration. This symmetry has recently been used to construct a new rotating vacuum geometry \cite{Barrientos:2025rjn}, which, surprisingly, is a fully regular geometry with no curvature singularity. 

We first examine the action of a Harrison transformation on the seed Ernst potentials, which generically takes the form
\begin{widetext}
\begin{equation}
\mathrm{H}_\beta:\left(\mathcal{E}_0, \Phi_0\right) \mapsto(\mathcal{E}, \Phi):=\left( \frac{\mathcal{E}_0}{1-2 \beta^* \Phi_0-|\beta|^2 \mathcal{E}_0}, \frac{\beta \mathcal{E}_0+\Phi_0}{1-2 \beta^* \Phi_0-|\beta|^2 \mathcal{E}_0}\right),
\end{equation}
\end{widetext}
where $\beta$ is a complex parameter whose real part is associated with the introduction of a magnetic field, while its imaginary part corresponds to an electric field. In the present construction, the symmetry is implemented in a magnetic WLP setting, so the transformation introduces the fields as external backgrounds rather than as fields sourced by localized charges. This follows from the asymptotic structure of the WLP spacetime under consideration, which is not asymptotically flat but asymptotically Levi-Civita \cite{Stephani:2003tm}.\footnote{The Levi-Civita spacetime can be represented by the line element
\begin{equation}
    \mathrm{d}s^2=-\rho^{4\sigma}\mathrm{d}t^2+\rho^{4\sigma(2\sigma-1)}(\mathrm{d}\rho^2+\mathrm{d}z^2)+\rho^{2(1-2\sigma)}\mathrm{d}\varphi^2,
\end{equation}
where $\sigma$ represents an effective gravitational mass per unit length of a cylindrical source.  
} Consequently, the electromagnetic configuration must preserve this non-trivial asymptotic behavior, a requirement incompatible with point-like charge sources. 
Note that this symmetry is capable of generating an electrovacuum solution from a purely vacuum seed.

For our purposes—namely, the magnetization of a static (magnetic) electrovacuum seed—it suffices to take the parameter $\beta$ to be real. Moreover, in this simplified setting, the effect of the symmetry can be described directly in the metric formalism, so the use of Ernst potentials is unnecessary. As shown in \cite{Dowker:1993bt}, the magnetizing transformation can be understood as follows: Owing to axisymmetry, the original metric can be decomposed as $\bar{g}_{\mu\nu}=(\bar{g}_{ij},\bar{g}_{\varphi\varphi})$. Assuming furthermore that the seed configuration is non-rotating, so that $\bar{g}_{i\varphi}=0$, and that only a seed magnetic field $A_{\varphi_0}$ is present, the resulting charged solution takes the form
\begin{subequations}
\label{chargesymmetry}
\begin{align}
\mathrm{d}s^2&=\Lambda^2\bar{g}_{ij}\mathrm{d}x^i\mathrm{d}x^j+\frac{\bar{g}_{\varphi\varphi}}{\Lambda^2}\mathrm{d}\varphi^2,\\
A_\varphi & =\Lambda^{-1}\left[A_{\varphi_0}+\frac{b}{2}\left( \bar{g}_{\varphi\varphi}+A_{\varphi_0}^2\right)\right], 
\end{align}\label{magnetizingsymmetry}
\end{subequations}
where
\begin{equation}
\Lambda  =\left(1+\frac{b}{2} A_{\varphi_0}\right)^2+\frac{b^2}{4} \bar{g}_{\varphi\varphi}.
\end{equation}
The parameter $b\sim\operatorname{Re}(\beta)$ modulates the intensity of the magnetic field. The rationale for employing this symmetry is straightforward. The two external fields—the original Bertotti–Robinson field present in the seed and the Melvin–Bonnor field introduced by the transformation—can be arranged to interact so that the total magnetic field vanishes, while the gravitational backreaction produced by these fields remains non-trivial. 

Now, let us introduce the azimuthal Inversion. This symmetry acts on a set of seed Ernst potentials $\{\mathcal{E}_0,\Phi_0\}$ as
\begin{equation}
\mathrm{I}:\left(\mathcal{E}_0, \Phi_0\right) \mapsto(\mathcal{E}, \Phi):=\left(\frac{1}{\mathcal{E}_0},\frac{\Phi_0}{\mathcal{E}_0}\right). \label{inversion}
\end{equation}
It can be obtained from an appropriate combination of Ehlers, dilation, and gravitational gauge transformations, taken in a specific limit of the associated parameters, which explains why it appears as a discrete symmetry \cite{Stephani:2003tm}. However, it has more recently been shown that no special limiting procedure is, in fact, required \cite{Barrientos:2023dlf,Barrientos:2024pkt,Barrientos:2025rjn}.

At this stage, the motivation for employing this symmetry becomes clear. When the seed gravitational and electromagnetic Ernst potentials are proportional, $\mathcal{E}_0 \sim \Phi_0$, the transformed magnetic potential becomes pure gauge, while the gravitational backreaction associated with the original magnetic configuration is preserved. This is precisely the mechanism at work when the azimuthal Inversion is applied to our seed configuration. 
At the level of the metric and gauge field components, the action of the azimuthal Inversion takes the form
\begin{equation}
    f=\frac{f_0}{f_0+A_{\varphi_0}^2}, \quad A_\varphi=-\frac{A_{\varphi_0}}{f_0+A_{\varphi_0}^2} . \label{invmetric}
\end{equation}
\vspace{0.2cm}

Having the two main ingredients at hand—the salient features of the accelerating Bertotti--Robinson black hole and the magnetizing Harrison and azimuthal Inversion symmetries—we now proceed to construct our novel static vacuum geometries.
\vspace{0.2cm}

\section{Novel Vacuum solutions}\label{sec3}

Let us now use the two symmetries explained above to construct two new families of vacuum solutions and disclose some of their main characteristics.

\subsection{Accelerating Bertotti--Robinson black hole in a Melvin--Bonnor Universe}

It is straightforward to apply the magnetizing transformation \eqref{magnetizingsymmetry} to embed the accelerating Bertotti–Robinson configuration \eqref{cmetricBR} into an external Melvin–Bonnor magnetic field. The resulting spacetime then assumes the form
\begin{align}
    \mathrm{d}s^{2}&=\frac{1}{\Omega^{2}}\left[\Lambda^{2}\left(-Q\mathrm{d}t^{2}+\frac{\mathrm{d}r^{2}}{Q}+\frac{r^{2}\mathrm{d}\theta^{2}}{P}\right)+\frac{r^{2}P\sin^{2}{\theta}}{\Lambda^{2}}\mathrm{d}\varphi^{2}\right],
\end{align}
where $\Omega$, $Q$, and $P$ retain the form given in \eqref{seedfunctions} and where the Melvin--Bonnor factor is given by 
\begin{widetext}
\begin{align}
    \Lambda&=\frac{-b\qty(B+\frac{b}{2})(B^{2}mr\cos^{2}{\theta}-\alpha r\cos{\theta}+1)+\qty(B^{2}+Bb+\frac{b^{2}}{2})\Omega}{B^{2}\Omega}.
\end{align}
The gauge field is given by the expression
\begin{align}
    A_{\varphi}&=-\frac{(B+b)[(-B^{2}mr\cos^{2}{\theta}+\alpha r\cos{\theta}-1)+\Omega]}{b(B^{2}mr\cos^{2}{\theta}-\alpha r\cos{\theta}+1)\qty(B+\frac{b}{2})-\Omega\qty(B^{2}+Bb+\frac{b^{2}}{2})}.
\end{align}
\end{widetext}
Here, $b$ represents the strength of the external Melvin–Bonnor magnetic field. Setting $b=0$ recovers the original seed spacetime, while taking $B=0$ yields the standard C-metric immersed in a Melvin–Bonnor field.
It is important to note that the magnetization procedure preserves the seed's conical defect: the interaction between $b$ and $B$ does not generate any effective force that could replace the role of the conical defect.
A point-like source structure for one of the fields is required to obtain such a result. In the non-accelerating case, the conical defect can be removed everywhere. 

It is straightforward to see that the two external fields, Bertotti–Robinson and Melvin–Bonnor, can interact in such a way that the total external electromagnetic field vanishes. This occurs when $b=-B$.\footnote{{Notice that the $\alpha=0$ case has already been constructed in \cite{Astorino:2025lih}, where a non-accelerating vacuum solution has been found.}} Importantly, this is not a generic feature; it arises specifically because the two fields possess distinct structural properties. A simple superposition of Melvin–Bonnor fields alone would not produce this effect. 

\begin{figure}[h]
        \includegraphics[width=8.5cm, height=8.5cm,left]{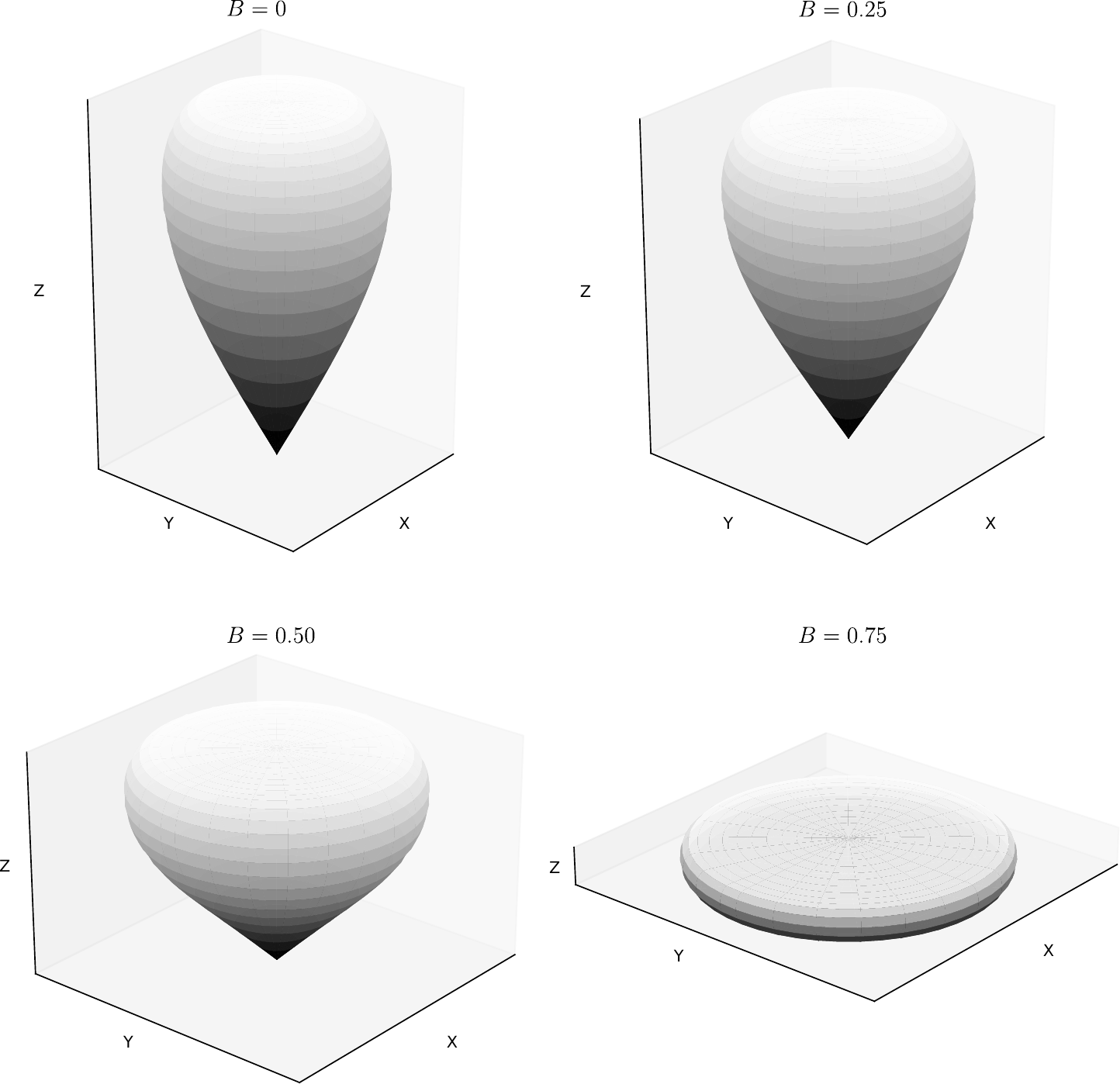}
    \caption{The horizon embedding for \eqref{vacuum1} is given for different values of the parameter $B$ and for a fixed acceleration $\alpha$. The higher the value of $B$, the more oblate the shape of the horizon.}
    \label{fig:embedding1}
\end{figure}

Our novel vacuum solution, written in a slightly different form than the electrovacuum configuration given in \eqref{cmetricBR}, is therefore represented by the line element 
\begin{widetext}
    \begin{equation}
\mathrm{d}s^2=\frac{\left(1-\alpha r\cos\theta+B^2 mr\cos^2\theta+\Omega \right)^2}{4\Omega^4}\left[-Q\mathrm{d}t^2+\frac{\mathrm{d}r^2}{Q}+\frac{r^2\mathrm{d}\theta^2}{P}\right]+\frac{4r^2P\sin^2\theta }{\left(1-\alpha r\cos\theta+B^2 mr\cos^2\theta+\Omega \right)^2}\mathrm{d}\varphi^2. \label{vacuum1}
\end{equation}
\end{widetext}
This metric is of algebraic type I. The locations of the horizons remain unchanged by the Harrison transformation and coincide with those of the seed spacetime. The special case $B=\alpha$ continues to appear, with more intricate consequences when $m=0$. While the horizon positions remain unaltered, their shapes are modified by the Melvin–Bonnor factor, producing a prolate or oblate deformation. The area of the horizon is given by 
\begin{equation}
    \mathcal{A}=\frac{16\pi m^2}{(1-2\alpha m+B^2m^2)^2(1+2\alpha m+B^2m^2)}, 
\end{equation}
and its deformation is evident in the embedding diagram in \autoref{fig:embedding1}.
A direct inspection of the Kretschmann invariant reveals a single curvature singularity located at $r=0$. Indeed, the limit $B=0$ recovers the vacuum C-metric. 

On the other hand, the background metric $(m=0)$ yields 
\begin{widetext}
    \begin{equation}
\mathrm{d}s^2=\frac{\left(1-\alpha r\cos\theta+\Omega_0 \right)^2}{4\Omega_0^4}\left[-(1+(B^2-\alpha^2)r^2)\mathrm{d}t^2+\frac{\mathrm{d}r^2}{(1+(B^2-\alpha^2)r^2)}+r^2\mathrm{d}\theta^2\right]+\frac{4r^2\sin^2\theta }{\left(1-\alpha r\cos\theta+\Omega_0 \right)^2}\mathrm{d}\varphi^2, 
\end{equation}
\end{widetext}
which turns out to be of Petrov type D. Here,  $\Omega_0^2=(1-\alpha r\cos\theta)^2+B^2r^2\sin^2\theta$. The conicity vanishes in this case, as seen by taking the zero-mass limit of the conicity in the black hole case.

\subsection{Inverted Accelerating Bertotti--Robinson black hole}

We now turn to the construction of an additional vacuum-accelerating solution, achieved by removing the external magnetic field via a different mechanism. Notably, for both types of seeds—the accelerating Bertotti–Robinson considered here and the Alekseev–García solution discussed in the Appendix \ref{appB}—the gravitational and electromagnetic Ernst potentials are proportional to each other, $\mathcal{E}_0 \propto \Phi_0$. Exploiting the azimuthal Inversion symmetry then naturally generates a vacuum solution without requiring any parameter tuning, while producing a spacetime with distinct geometric properties. Using the expressions \eqref{invmetric} for the seed spacetime written in magnetic WLP form 
\begin{align}
\mathrm{d}s_0^{2}&=f_{0}\mathrm{d}\varphi^{2}+\frac{1}{f_{0}}\qty[e^{2\gamma_{0}}(\mathrm{d}\rho^{2}+\mathrm{d}z^{2})-\rho^{2}\mathrm{d}t^{2}],\\
    A_\varphi&=A_{\varphi0} \mathrm{d}\varphi,
\end{align}
where 
\begin{align}
    f_{0}&=\frac{r^{2}P\sin^{2}{\theta}}{\Omega^{2}},\\
    \rho&=\frac{r\sin{\theta}\sqrt{PQ}}{\Omega^{2}},\\
    A_{\varphi_{0}}&=\frac{1}{B}(r\partial_{r}\Omega-\Omega+1),
\end{align}
allows for constructing the new vacuum solution
\begin{align}
\mathrm{d}s^{2}&=f\mathrm{d}\varphi^{2}+\frac{r^{2}P\sin^{2}{\theta}}{f\Omega^{4}}\qty[-Q\mathrm{d}t^{2}+\frac{\mathrm{d}r^{2}}{Q}+\frac{r^{2}\mathrm{d}\theta^{2}}{P}], \label{vacuum2.1}
\end{align}
where 
\begin{align}
    f&=-\frac{B^{2}(1+B^{2}mr\cos^{2}{\theta}-\alpha r\cos{\theta}+\Omega)} {4(1+B^{2}mr\cos^{2}{\theta}-\alpha r\cos{\theta}-\Omega)},\label{vacuum2.2}\\
    A_{\varphi}&=-\frac{B}{2}.
\end{align}
The magnetic potential becomes a constant that can be removed by a gauge transformation. 
We recall that the non-Killing sector of the metric, as encoded by the function $\gamma$, remains unchanged. 

Notice that the non-accelerating sub-case $\alpha=0$ of this line element has not been introduced in the literature before, which is why we believe that most of the analysis can be carried out for such a simpler scenario. As we will see below, this case is very appealing because the spacetime does not contain the usual curvature singularity along the symmetry axis associated with solutions affected by an azimuthal Inversion, and it also does not exhibit any conical deficit. 
We analyze the $\alpha=0$ hereafter. 

{The locations of the black hole horizons remain unchanged under the transformation and thus coincide with those of the seed spacetime.
The induced metric on a spatial cross-section of the horizon is
\begin{equation}
\mathrm{d}s_h^2
=
f_h\,\mathrm{d}\varphi^2
+
\frac{r_h^4\sin^2\theta}
     {f_h\Omega_h^4}
\,\mathrm{d}\theta^2,
\end{equation}
where
\begin{equation}
f_h=
\frac{1+B^2m^2\cos^2\theta}
     {4m^2\sin^2\theta},
\qquad
\Omega_h=
\frac{1+B^2m^2}
     {1-B^2m^2}.
\end{equation}
Moreover, near $\theta=0$ one finds
\begin{equation}
g_{\varphi\varphi}
\sim
\frac{1+B^2m^2}{4m^2\theta^2},
\qquad
g_{\theta\theta}
\sim
\frac{64m^6}{(1+B^2m^2)^5}\theta^4,
\end{equation}
and similarly near $\theta=\pi$. Consequently, the azimuthal orbits do not collapse at the endpoints, in contrast with the behavior of a regular spherical pole. Instead, $g_{\varphi\varphi}$ diverges, and the horizon exhibits the same type of singular end structure found in the Schwarzschild--Levi-Civita solution. Removing these singular points, the regular part of the horizon is diffeomorphic to
\begin{equation*}
(0,\pi)\times S^1 \cong \mathbb R\times S^1,
\end{equation*}
and therefore possesses cylindrical topology.
An embedding diagram of the horizon is shown in \autoref{fig:embedding2}.
 
\begin{figure}[h]
    \centering    \includegraphics[scale=0.5]{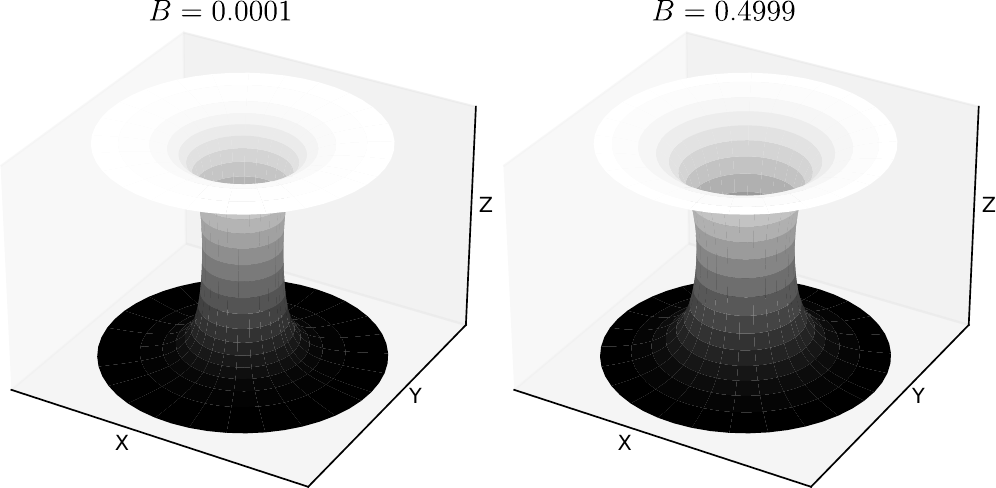}
    \caption{{The horizon embedding for \eqref{vacuum2.1} in the case of vanishing acceleration
    is given for different values of the parameter $B$ and for $m=2$. As expected, the structure of the horizon follows that of the Schwarzschild--Levi-Civita black hole.}}
    \label{fig:embedding2}
\end{figure}

On the other hand, the conical defect is found to be
\vspace{0.2cm}
\begin{equation}
     \lim_{\theta\rightarrow 0,\pi}\frac{1}{\sin\theta}\int_0^{2\pi}\sqrt{\frac{g_{\varphi\varphi}}{g_{\theta\theta}}}\mathrm{d}\varphi= 2\pi\frac{B^2m^2 +1 }{16}, \label{conicity2ndcase}
\end{equation}
which can be removed everywhere. This follows from the analysis of the non-accelerating case: the Inversion symmetry does not introduce any conical singularity into the original Schwarzschild–Bertotti–Robinson metric. 

The Kretschmann invariant signals the presence of a curvature singularity only at $r=0$, proving that the symmetry axis is entirely regular. Additionally, algebraically, the geometry remains of Petrov type I.

In contrast to the previous case, taking the limit $B \to 0$, after suitable rearrangements in the metric, yields the Schwarzschild–Levi-Civita spacetime
\begin{widetext}
\begin{equation}
     \mathrm{d}s^{2}=r^4\sin^4\theta\left[-\left(1-\frac{2m}{r}\right)\mathrm{d}t^{2}+\frac{\mathrm{d}r^{2}}{\left(1-\frac{2m}{r}\right)}+r^{2}\mathrm{d}\theta^2\right]
  +\frac{\mathrm{d}\varphi^{2}}{r^2\sin^2\theta},
\end{equation}
\end{widetext}
as expected from the application of the azimuthal Inversion \cite{Barrientos:2024uuq}.

For this non-accelerating case, the following comments on the thermodynamics of the spacetime are in order. {The black hole temperature can be computed, as the surface gravity $\kappa_s$ on the event horizon, via the standard relation (see, e.g., \cite{Wald:1984rg})}
\begin{equation}
T=\frac{\kappa_s}{2\pi}=\frac{1}{2\pi}\sqrt{-\frac{1}{2}\xi_{\mu;\nu}\xi^{\mu;\nu}}=\frac{(1+B^2m^2)^2}{8\pi m},
\end{equation}
where we evaluate $\xi=\partial_t$ at the event horizon $r=r_h$. The mass of the black hole is calculated via the Komar integral 
\begin{equation}
    M=-\frac{1}{8\pi}\oint_{S_\infty}\nabla^\mu\xi^\nu n_{[\mu}u_{\nu]}\mathrm{d}S_{\infty}=\frac{16m}{1+B^2m^2}, 
\end{equation}
with $\mathrm{d}S_{\infty}=\sqrt{g_{\theta\theta}g_{\varphi\varphi}}\mathrm{d}\theta \mathrm{d}\varphi$ and where the timelike and spacelike orthonormal vectors to $S_\infty$, $u$ and $n$, satisfy $u^\mu u_\mu=-1$, $n^\mu n_\mu=1$ and $u^\mu n_\mu=0$.

The entropy follows the area law, which, after the elimination of the conicity \eqref{conicity2ndcase}, yields 
\begin{equation}
S=\frac{\mathcal{A}}{4}=\frac{64\pi m^2}{\left(1+B^2m^2\right)^3}.
\end{equation}
These thermodynamic quantities satisfy the first law of black hole thermodynamics
\begin{equation}
\delta M=T\delta S,
\end{equation}
provided the Killing vector $\xi$ is renormalized according to  $\bar{\xi}=\frac{\xi}{4\sqrt{1+B^2m^2}}$. With this normalization, the physical mass and temperature become
\begin{equation}
 \bar{M}=\frac{4m}{\left(1+B^2m^2\right)^{3/2}},\qquad \bar{T}=\frac{(1+B^2m^2)^{3/2}}{32\pi m}.
\end{equation}
These new quantities also satisfy the Smarr relation
\begin{equation}
\bar{M}=2\bar{T}S,
\end{equation}
and its quadratic generalization
\begin{equation}
\bar{M}^2=\frac{S}{4\pi}.
\end{equation}
Finally, via the relation $S=\frac{1}{16\pi \bar{T}^2}$, it can be observed that the specific heat of the solution is
\begin{equation}
C=\bar{T}\frac{\partial S}{\partial \bar{T}}=-2S,
\end{equation}
which is always negative. The solution is therefore locally thermodynamically unstable, in analogy with the standard Schwarzschild case. 

Now, the background geometry yields
\begin{widetext}
\begin{equation}
     \mathrm{d}s^{2}=\frac{4r^2\sin^2\theta(\Omega_0-1)}{B^2\Omega_0^{4}(\Omega_0+1)}\qty[-(1+B^2r^2)\mathrm{d}t^{2}+\frac{\mathrm{d}r^{2}}{1+B^2r^2}+r^{2}\mathrm{d}\theta^2]
  +\frac{B^{2}}{4}\frac{(\Omega_0+1)}{(\Omega_0-1)}\mathrm{d}\varphi^{2},
\end{equation}
\end{widetext}
where here $\Omega_0^2=1+B^2r^2\sin^2\theta$, which is of Petrov algebraic type D. Indeed, the limit around $B=0$ precisely reproduces the Levi-Civita spacetime with Levi-Civita parameter $\sigma=1$, which is known to be of type D as well. For a non-trivial value of $B$, the large-$r$ limit is not particularly illuminating and leads to a cumbersome expression; again, when expanded around $B=0$, it reduces to a Levi-Civita spacetime with $\sigma=1$.
\section{Final remarks}\label{sec4}

The aim of this paper has been to present two strategies for enlarging the spectrum of vacuum black hole solutions in GR. Although such a task may initially appear unlikely to succeed, our results suggest that exploring algebraically general geometries offers a promising avenue for further progress.

We have demonstrated that new algebraically general static vacuum solutions can be generated by embedding accelerating Bertotti–Robinson seeds into external electromagnetic backgrounds and subsequently removing the net electromagnetic field while preserving the gravitational backreaction, and, separately, by using the Inversion symmetry with no further parameter tuning.  

For those solutions obtained via the Harrison symmetry, magnetization followed by the choice $b=-B$ yields the vacuum geometry described by \eqref{vacuum1}, where the electromagnetic sector vanishes identically while the metric retains a non-trivial magnetization factor. These spacetimes are generically Petrov type I, possess horizons located at the same radial positions as in the seed solution, and exhibit a single curvature singularity at $r=0$, as indicated by the Kretschmann scalar. The geometry of the horizons is modified by the magnetizing function, producing deformations without altering their coordinate locations. A particularly interesting regime arises when $B=\alpha$, for which the acceleration horizon disappears.

For the solutions constructed with the azimuthal Inversion map, the azimuthal transformation produces the vacuum metric \eqref{vacuum2.1}-\eqref{vacuum2.2}, where the magnetic potential reduces to a pure gauge term that can be removed by a trivial redefinition. No tuning of parameters is required. The resulting spacetime is again algebraically general, while the horizon structure remains unchanged. 

In the non-accelerating limit $\alpha=0$, the solution becomes particularly appealing: the conical defect can be completely removed, and the symmetry axis is free of any curvature singularity. This feature contrasts with other static vacuum geometries obtained via azimuthal Inversion symmetry, such as the Schwarzschild–Levi-Civita geometry \cite{Barrientos:2024uuq}, which corresponds to the $B=0$ limit of our new vacuum configuration. The regularization of the axis appears to be a remnant of the initial external magnetic field, whose imprint in the line element is not eliminated by the transformation.

{At this stage, it is important to emphasize the utility of the Inversion transformation and the genuinely non-trivial effects that it may produce.

Beyond the solutions presented in this work, this symmetry has also been employed to construct the Kerr--Levi-Civita geometry \cite{Barrientos:2025rjn} and its static counterpart, the Schwarzschild--Levi-Civita spacetime \cite{Barrientos:2024uuq}.\footnote{Although this solution arises through the use of a magnetic Inversion symmetry, it has often been identified via Buchdahl symmetry \cite{Buchdahl:1956zz}, which effectively provides a formulation of Inversion symmetry directly at the level of the metric.}

In a recent paper \cite{Astorino:2026kuv}, published after the first version of the present work had been completed, it was argued that these geometries can be recovered as singular limits of the so-called swirling solutions, which are generated by applying Ehlers transformations to a given electrovacuum seed metric.\footnote{It is worth noting that, contrary to the claim in \cite{Astorino:2022aam}, swirling spacetimes were originally constructed in \cite{Contopoulos:2015wra}, seven years earlier, without proper acknowledgment in subsequent works.} By this, it is meant that a suitable combination of coordinate transformations and parameter redefinitions allows one to recover members of the Levi-Civita family from the corresponding family of swirling spacetimes.

A particularly instructive example is provided by the relation between the Schwarzschild--swirling spacetime and the static Schwarzschild--Levi-Civita geometry. In Eq.~(A5) of Ref.~\cite{Astorino:2026kuv}, a set of coordinate transformations and parameter redefinitions is introduced such that, in the singular limit $j\rightarrow\infty$ ($j$ being the swirling parameter), the Schwarzschild--Levi-Civita line element is recovered from the Schwarzschild--swirling metric. We fully agree with this result, which establishes the local equivalence between the geometry obtained through this limiting procedure and the Schwarzschild--Levi-Civita solution generated by the Inversion symmetry.

The remaining question is whether this singular limit also establishes \emph{global} equivalence. Singular limits relating different exact solutions are abundant in the literature, and they often reveal non-trivial connections between apparently distinct geometries. However, they do not necessarily preserve the global structure of the spacetime. The issue, therefore, is not only whether the local limit exists—it clearly does—but whether the resulting spacetime is globally identical to the Schwarzschild--Levi-Civita solution. Our present understanding is that this is not the case. Consequently, we do not agree with the statement that the existence of the singular limit alone renders the use of the Inversion symmetry redundant. To clarify this point, let us examine the possible ways of recovering the Schwarzschild--Levi-Civita geometry starting from the Schwarzschild--swirling family.

\begin{enumerate}[label=\roman*.]
\item Start from the Schwarzschild--swirling spacetime, whose azimuthal coordinate has a period $2\pi$. Apply the coordinate transformations proposed in Ref.~\cite{Astorino:2026kuv} and subsequently take the limit. The resulting metric is locally identical to the Schwarzschild--Levi-Civita geometry. However, the inherited period of the new azimuthal coordinate, here denoted by $\varphi$, shrinks to zero, so that the proper length of the orbits generated by $\partial_\varphi$ vanishes. Consequently, local equivalence is achieved, whereas global equivalence is not.

\item Start from a spacetime having the same local form as Schwarzschild--swirling, but with the azimuthal coordinate identified with a period $2\pi J^2$, where $J$ here denotes the redefined parameter introduced in Eq.~(A5) of Ref.~\cite{Astorino:2026kuv}. After applying the same coordinate transformations and taking the limit, one indeed obtains a spacetime that is both locally and globally equivalent to Schwarzschild--Levi-Civita. The crucial point, however, is that the initial geometry is no longer the Schwarzschild--swirling spacetime itself, but rather a globally different spacetime sharing the same local metric. In other words, the seed has been modified.

\item Start from the Schwarzschild--swirling spacetime with azimuthal period $2\pi$, and first apply a rescaling transformation in potential space. It is well known that such a transformation is pure gauge when the associated Killing coordinate is non-compact. However, when the Killing coordinate is compact, the situation is different. 
The transformed solution may be brought back to the local form of the original metric through a rescaling of the Killing coordinate together with a Weyl rescaling of the orbit-space metric, the latter always being a symmetry of the reduced field equations. By choosing the rescaled azimuthal period to be $2\pi J^2$, one precisely recovers the starting point described in case ii), after which the same reasoning applies.
\end{enumerate}

Cases i) and ii) clearly illustrate the two possible outcomes. Either one starts from the Schwarzschild--swirling spacetime and reaches a geometry that is only locally equivalent to Schwarzschild--Levi-Civita, or one starts from a globally different seed and obtains the Schwarzschild--Levi-Civita spacetime both locally and globally.

Case iii) is particularly interesting because it allows one to begin with the Schwarzschild--swirling spacetime and ultimately recover the Schwarzschild--Levi-Civita geometry, both locally and globally. However, it is essential to recognize that this construction does not rely solely on the coordinate transformations and singular limit proposed in Ref.~\cite{Astorino:2026kuv}. It additionally requires a rescaling symmetry acting on the potentials, which is not pure gauge when the associated Killing direction is compact. A detailed analysis of this last statement will be presented in \cite{SoonAdolfo}.

Therefore, case iii) does not establish that the Schwarzschild--Levi-Civita spacetime is globally contained within the Schwarzschild--swirling family through coordinate transformations and a singular limit alone. Rather, it requires an additional non-trivial symmetry transformation. This distinction constitutes the essential reason why we assert that the Inversion symmetry remains a genuinely useful and non-redundant solution-generating technique when applied along an azimuthal Killing direction, which is naturally compact in the axisymmetric spacetimes considered here.

It is important to emphasize that we are not attributing any global effect to the Inversion symmetry itself. Like the hidden symmetry transformations from which it originates, the Inversion symmetry acts locally. Therefore, if one starts from a Schwarzschild seed with a given identification of the azimuthal coordinate, the resulting Schwarzschild--Levi-Civita geometry inherits precisely the same identification. The global properties are thus inherited from the seed spacetime, not generated by the Inversion transformation.

From the perspective of symmetry transformations, it is clear that no particular transformation—whether Ehlers or Inversion—should be regarded as more fundamental than another. All such transformations arise from the action of group elements, and due to the transitive nature of the group action, any transformation can be expressed as a composition of others. In particular, Inversion, as well as the potential transformation discussed in \cite{Clement:1997tx}, is realized through the action of elements belonging to a compact subgroup of $SU(2,1)$ isomorphic to $SO(2)$, corresponding to rotations by fixed angles. Importantly, the group element implementing Inversion does not belong to the one-parameter Ehlers subgroup. This explains why recovering Inversion from an Ehlers transformation requires both additional transformations and a limiting procedure.

The so-called enhanced Ehlers transformation, which is an Ehlers transformation augmented with an ad hoc gauge symmetry transformation (with gauge symmetry parameter fixed) and that reduces to Inversion in the limit $j \to \infty$, does not, in general, produce the same physical effects as a standard Ehlers transformation. Rather, it consists of an Ehlers transformation, followed by a duality/rescaling transformation, and a shift in the gravitational twist scalar. While the shift of the twist scalar corresponds to a gauge freedom in the space of potentials and does not affect the spacetime metric, the duality/rescaling transformation is not generically a gauge symmetry.
Indeed, one can show that the local form of the metric---when expressed in coordinates adapted to the Killing vector defining the Ernst potentials---remains invariant under a duality/rescaling transformation, provided this is accompanied by a global Weyl rescaling of the orbit-space metric (which itself is a symmetry of the reduced field equations) and a corresponding rescaling of the Killing coordinate. If the Killing coordinate is non-compact, the spacetime metric remains unchanged, and the transformation acts non-trivially only on the Maxwell field, if included. In the absence of a gauge field, the transformation is therefore effectively pure gauge. However, if the Killing coordinate is compact and periodic, the situation changes: although the metric functions remain invariant, the periodicity of the transformed Killing coordinate differs from the original one. Enforcing the original periodicity amounts to altering the global identification, thereby leading to a physically distinct spacetime manifold. This is precisely the situation in the present case, as both the swirling geometries and the Inversion transformations considered here are defined with respect to the azimuthal Killing vector, whose orbits are naturally periodic and therefore intrinsically identified.
Therefore, even if a singular limit applied to a swirling geometry yields a metric that is locally equivalent to one obtained via Inversion symmetry, the resulting spacetimes are globally distinct.}

Two additional results are presented in the appendices. First, we construct a vortex-like generalization of the vacuum metric obtained through the Harrison transformation. Second, we present two additional novel vacuum geometries derived from the Alekseev–Garc\'ia black hole as the seed, and we leave their detailed analysis for future work. By explicit construction, we show that the interaction of the external fields can mutually cancel, and that Inversion symmetry can again remove them.

These constructions indicate that the spectrum of vacuum solutions in GR may still be enlarged. In the present work, we have employed only seed configurations in which the black hole embedded in the external field carries no point-like charge. In \cite{Ovcharenko:2026byw}, Reissner--Nordström configurations on a Bertotti--Robinson background were also obtained. It would therefore be interesting to apply the techniques developed here to search for additional vacuum metrics and investigate their geometric properties and particle dynamics. 
\acknowledgments
The authors thank Konstantinos Pallikaris and Mokhtar Hassaine for insightful discussions and comments on the novelty of Inversion symmetry and related topics.
J.B. acknowledges the support of ANID through grant FOVI250117 and the hospitality of the Institute of Mathematics, Czech Academy of Sciences (RVO 67985840). A.C. is partially supported by FONDECYT grant 1250318. The work of K.M. is funded by Beca Nacional de Doctorado ANID grant No. 21231943.

\appendix
\section{Novel vortex-like vacuum solutions}\label{appA}

Stationary generalizations of the vortex-like version of our vacuum metrics can, in principle, be constructed straightforwardly, although the procedure is computationally demanding. Here, we present the line element corresponding to a vortex-like extension of our vacuum solution \eqref{vacuum1}, leaving the rotational function implicitly defined through a twist equation to be solved. It reads 
\begin{equation}
    \mathrm{d}s^2=f(\mathrm{d}\varphi+\omega \mathrm{d}t)^2+\frac{1}{f}[e^{2\gamma_{0}}(\mathrm{d}\rho^{2}+\mathrm{d}z^{2})-\rho^{2}\mathrm{d}t^{2}],
\end{equation}
where $\gamma_0$ is obtained by expressing the seed metric in magnetic WLP form and remains unchanged under the transformation,\footnote{It is known that vortex-like solutions can be readily generated via magnetic Ehlers transformations, which is why we do not present the explicit construction.} and where
\begin{widetext}
    \begin{equation}
\begin{aligned}
f &=
\frac{
4 r^2 P\sin^2\theta
}{
\left(1 - \alpha r \cos\theta + B^2 m r \cos^2\theta + \Omega\right)^2
\left(
1 + \dfrac{16 j^2 r^4 P^2\sin^4\theta}
{\left(1 - \alpha r \cos\theta + B^2 m r \cos^2\theta + \Omega\right)^4}
\right)},\\
h&=
\frac{
16j r^4 P^2\sin^4\theta 
}{
\left(1 - \alpha r \cos\theta + B^2 m r \cos^2\theta + \Omega\right)^4
\left(
1 + \dfrac{16 j^2 r^4 P^2\sin^4\theta }
{\left(1 - \alpha r \cos\theta + B^2 m r \cos^2\theta + \Omega\right)^4}
\right)}.
\end{aligned}
\end{equation}
\end{widetext}
The function $h$ is a twist potential that, via the following twist equation components 
\begin{equation}
\partial_\theta \omega
-
\frac{Q r^2 \sin\theta \partial_r h}
{f^2 \Omega^2}=0,\qquad \partial_r \omega
+
\frac{P\sin\theta \partial_\theta h}
{f^2 \Omega^2}=0.
\end{equation}
allows the integration of the rotation function $\omega$.
\vspace{0.2cm}
\\
\section{New vacuum geometries starting from Alekseev-Garc\'ia configuration}\label{appB}

The Alekseev–Garc\'ia black hole line element, using a coordinate system $\{t,\rho,z,\varphi\}$, where the Killing directions $\{t,\varphi\}$ retain their usual interpretation, and the remaining non-Killing coordinates must be specified carefully, as they do not follow their standard canonical roles, can be written as
\begin{widetext}
\begin{equation}
\mathrm{d} s^2=-e^{2 \psi} \cosh ^2 \frac{z}{B} \mathrm{d} t^2+e^{2 \gamma}\left(\mathrm{d} \rho^2+\mathrm{d} z^2\right)+e^{-2 \psi} B^2 \sin ^2 \frac{\rho}{B} \mathrm{d} \varphi^2, \label{solAG}
\end{equation}
where the metric functions $\{\psi,\gamma\}$ are given by 
\begin{align}
e^{2 \psi}=\frac{\left(\mathcal{R}_{+}+\mathcal{R}_{-}-2 m \cos \frac{\rho}{B}\right)^2}{\left(\mathcal{R}_{+}+\mathcal{R}_{-}\right)^2-4 m^2},\quad e^{2 \gamma}=\frac{\left(\mathcal{R}_{+}+\mathcal{R}_{-}-2 m \cos \frac{\rho}{B}\right)^2}{4 \mathcal{R}_{+} \mathcal{R}_{-}}\left[\frac{\mathcal{R}_{+}-B \sinh \frac{z}{B}+(l+m) \cos \frac{\rho}{B}}{\mathcal{R}_{-}-B \sinh \frac{z}{B}+(l-m) \cos \frac{\rho}{B}}\right]^2. \label{seedfunctionsAG}
\end{align}
\end{widetext}
Here, it has been defined 
\begin{equation}
\mathcal{R}_{ \pm}^2=\left(l \pm m-B \sinh \frac{z}{B} \cos \frac{\rho}{B}\right)^2+B^2 \cosh ^2 \frac{z}{B} \sin ^2 \frac{\rho}{B}, \label{seedrod}
\end{equation}
where $\{m,l,B\}$ is a set of constants parametrizing the solution. The line element is complemented by the magnetic gauge potential, fixed up to an arbitrary constant duality rotation, 
\begin{equation}
A=-B\frac{\mathcal{R}_{+}+\mathcal{R}_{-}+2 m}{\mathcal{R}_{+}+\mathcal{R}_{-}-2 m \cos \frac{\rho}{B}}\left(1-\cos \frac{\rho}{B}\right)\mathrm{d}\varphi.
\end{equation}

This electrovacuum configuration can be interpreted as describing a Schwarzschild black hole immersed in a homogeneous magnetic and gravitational background of the Bertotti–Robinson type. Indeed, in the limit $m=0$, the solution exactly reduces to the Bertotti–Robinson universe. 

The topology of the Bertotti--Robinson background plays a crucial role in understanding the geometric properties of this highly non-trivial electrovacuum solution. In particular, spatial sections of the Bertotti–Robinson metric have $\mathbb{R}\times S^{2}$ topology, and as a consequence, the Weyl–Papapetrou radial coordinate no longer ranges over the standard interval $0\leq \rho < \infty$. Instead, it is restricted to $0\leq \rho \leq \pi B$, reflecting the emergence of a second set of fixed points of the azimuthal Killing vector $\partial_{\varphi}$, corresponding to a second axis of symmetry located at the antipodal value $\rho=\pi B$.

Although we do not pursue a detailed geometric analysis of the solutions here, we show by explicit construction that two novel vacuum spacetimes can be generated from the Alekseev–García configuration, in direct analogy with the procedure applied to \eqref{cmetricBR}. The first solution is obtained via standard magnetization and, after suitably tuning the parameters $B$ and $b$, it reads 
\begin{widetext}
\begin{align}
\mathrm{d} s^2&=\Lambda^2\left[-e^{2 \psi} \cosh ^2 \frac{z}{B} \mathrm{~d} t^2+e^{2 \gamma}\left(\mathrm{~d} \rho^2+\mathrm{d} z^2\right)\right]+\frac{e^{-2 \psi} B^2 \sin ^2 \frac{\rho}{B}}{\Lambda^2} \mathrm{~d} \varphi^2\label{magentizedAGsolmetric},
\end{align}
\end{widetext}
where 
\begin{equation}
    \Lambda=\frac{1}{2}\frac{(\mathcal{R}_+ + \mathcal{R}_- - 2m)(1+\cos\frac{\rho}{B})}{(\mathcal{R}_+ + \mathcal{R}_- - 2m\cos\frac{\rho}{B})}. \label{vacuumnewAG}
\end{equation}
On the other hand, the solution obtained via the azimuthal Inversion yields 
\begin{equation}
    \mathrm{d}s^2= \frac{1}{f}\left[-B^2\cosh^2\frac{z}{B}\sin^2\frac{\rho}{B}\mathrm{d}t^2+e^{2\bar{\gamma}}(\mathrm{d}\rho^2+\mathrm{d}z^2)\right]+f\mathrm{d}\varphi^2,
\end{equation}
where $\bar{\gamma}=\gamma+\psi$, with $\{\gamma,\psi\}$ as defined in \eqref{seedfunctionsAG} and where 
\begin{equation}
    f=\frac{1}{4B^2}\frac{(\mathcal{R}_{+} + \mathcal{R}_{-} - 2m)(1+\cos\frac{\rho}{B})}{(\mathcal{R}_{+} + \mathcal{R}_{-} + 2m)(1-\cos\frac{\rho}{B})}.
\end{equation}
The quantity $\mathcal{R}_{\pm}$ retains its definition as given in \eqref{seedrod}.
The transformed magnetic field becomes a pure gauge, $A=\frac{1}{2B}\mathrm{d}\varphi$.

Note that in both solutions we have considered the case $l=0$.

\bibliography{apssamp}

\end{document}